\documentclass[prx,superscriptaddress]{revtex4-2}
\usepackage[utf8]{inputenc}
\usepackage{color}
\usepackage{amsmath,amssymb}
\usepackage{graphicx}
 
\begin{document}
\title{Revealing domain wall stability during ultrafast demagnetization}

\author{Hung-Tzu Chang}
\thanks{H. -T. C. and S. Z. contributed equally to this work.}
\affiliation{Max Planck Institute for Multidisciplinary Sciences, 37077 G\"ottingen,
Germany}

\author{Sergey Zayko}
\thanks{H. -T. C. and S. Z. contributed equally to this work.}
\email{szayko@gwdg.de}
\affiliation{Max Planck Institute for Multidisciplinary Sciences, 37077 G\"ottingen,
Germany}

\author{Timo Schmidt}
\affiliation{Institute of Physics, University of Augsburg, 86135 Augsburg, Germany }

\author{Ofer Kfir}
\affiliation{School of Electrical Engineering, 
Tel Aviv University, Tel Aviv 69978, Israel}

\author{Murat Sivis}
\affiliation{Max Planck Institute for Multidisciplinary Sciences, 37077 G\"ottingen,
Germany}
\affiliation{4\textsuperscript{th} Physical Institute, University of G\"ottingen,
37077 G\"ottingen, Germany}

\author{Johan H. Mentink}
\affiliation{Radboud University, Institute for Molecules and Materials, Nijmegen, 6525 AJ, Netherlands}

\author{Manfred Albrecht}
\affiliation{Institute of Physics, University of Augsburg, 86135 Augsburg, Germany }

\author{Claus Ropers}
\email{claus.ropers@mpinat.mpg.de}
\affiliation{Max Planck Institute for Multidisciplinary Sciences, 37077 G\"ottingen,
Germany}
\affiliation{4\textsuperscript{th} Physical Institute, University of G\"ottingen,
37077 G\"ottingen, Germany}

\begin{abstract}
The ultrafast control of nanoscale spin textures such as magnetic domain walls or skyrmions is essential for advancing high-speed, high-density spintronics. However, imaging their dynamics will require a technique that combines nanometer spatial and femtosecond temporal resolution. Introducing ultrafast sub-wavelength imaging in the extreme ultraviolet, we track domain wall properties during ultrafast demagnetization in ferro- and ferrimagnetic thin films. We reveal that domain walls remain invariant in position, shape, and width, down to a demonstrated sub-nanometer precision, for up to 50\% demagnetization. Stronger excitation causes stochastic nanoscale domain switching. This previously unobservable robustness of laser-excited domain walls highlights the localized nature of photoinduced demagnetization and presents both challenges and opportunities for all-optical magnetic control. The presented technique can be generalized to directly probe nanoscale dynamics in spintronic materials and devices.
\end{abstract}
\maketitle

The manipulation of nanoscale spin textures such as magnetic domain walls (DW) with femtosecond laser pulses can facilitate data processing at ultrafast speeds, and underpin the development of next-generation information storage \cite{parkinMagneticDomainWallRacetrack2008,fertMagneticSkyrmionsAdvances2017,kimelWritingMagneticMemory2019}. Therein, photoinduced demagnetization plays an essential role \cite{beaurepaireUltrafastSpinDynamics1996,Kirilyuk2010}. Nevertheless, the interplay between electronic, structural, and spin degrees of freedom, and inhomogeneity from nanoscale domain patterns, render the mechanisms of demagnetization and DW dynamics a challenging subject under ongoing debate. Among the manifold phenomena in ultrafast magnetism, a wide variety of processes and mechanisms such as energy transfer to the spin system \cite{beaurepaireUltrafastSpinDynamics1996}, loss of angular momentum due to spin-flip scattering and phonon excitation \cite{koopmansExplainingParadoxicalDiversity2010,dornesUltrafastEinsteinHaas2019,tauchertPolarizedPhononsCarry2022}, as well as non-local spin transport \cite{battiatoSuperdiffusiveSpinTransport2010,melnikovUltrafastTransportLaserExcited2011}, have been observed or proposed.  For example, optical microscopy measurements on DW dynamics during magneto-optical switching showed domain wall movement with velocities up to $\sim 1$ km/s and stochastic nucleation of magnetic domains \cite{medapalliMultiscaleDynamicsHelicitydependent2017,quessabResolvingRoleMagnetic2019,prabhakaraStudyDomainWall2022,khusyainovLaserinducedHelicityTexturedependent2024}. Such phenomena can be explained by local heating and are consistent with experimental observations for skyrmion nucleation \cite{koshibaeCreationSkyrmionsAntiskyrmions2014,jeCreationMagneticSkyrmion2018,buttnerObservationFluctuationmediatedPicosecond2021,jugeSkyrmionsSyntheticAntiferromagnets2022}. On the other hand, time-resolved studies employing extreme ultraviolet (XUV) or soft X-ray scattering suggested significantly faster DW velocities reaching $\sim$70 km/s \cite{hennesLaserinducedUltrafastDemagnetization2020,fanUltrafastMagneticScattering2022,zhouhagstromSymmetrydependentUltrafastManipulation2022,zusinUltrafastPerturbationMagnetic2022,jangidExtremeDomainWall2023}, far exceeding the Walker breakdown velocities at which the DW structure becomes unstable \cite{schryer1974}. Moreover, DW broadenings up to $\sim$50\% of their original widths, attributed to a decrease in uniaxial anisotropy or non-local superdiffusive spin transport \cite{pfauUltrafastOpticalDemagnetization2012,zusinUltrafastPerturbationMagnetic2022,hennesLaserinducedUltrafastDemagnetization2020}, were inferred from ultrafast diffractive probing. The broad range of both experimental observations and theoretical explanations not only illustrate the richness of non-equilibrium magnetic phenomena, but also the challenges involved in developing a comprehensive approach to inspect and elucidate processes in ultrafast magnetism. Experimentally, optical microscopy can track local magnetization dynamics but typically lacks the spatial resolution to resolve magnetic domain walls. Scattering experiments in reciprocal space enable extraction of nanoscale statistical properties of domain walls. However, this inevitably relies on assumptions on the domain morphology and DW profiles, as discussed in, e.g. Refs. \cite{zusinUltrafastPerturbationMagnetic2022,suturinShortNanometerRange2023}, which motivates ultrafast real-space imaging at sufficient resolution.

Photoinduced magnetization dynamics typically unfold on sub-picosecond times and lengths below $20$~nm. Time-resolved magnetic imaging experiments can reach such scales, but thus far only separately.  Figure~1a summarizes the literature on DW dynamics inferred from XUV and X-ray scattering studies (red-shaded region)  \cite{pfauUltrafastOpticalDemagnetization2012,kerberFasterChiralCollinear2020,hennesLaserinducedUltrafastDemagnetization2020,leveilleUltrafastTimeevolutionChiral2022,fanUltrafastMagneticScattering2022,zusinUltrafastPerturbationMagnetic2022,zhouhagstromSymmetrydependentUltrafastManipulation2022,jangidExtremeDomainWall2023}, together with the current state of the art in magnetic imaging (blue region). Time-resolved microscopy with soft-X-ray synchrotron radiation achieves a spatial resolution down to 10~nm \cite{girardiThreedimensionalSpinwaveDynamics2024,butcher2024a,donnelly2020,weigand2022a, buttnerDynamicsInertiaSkyrmionic2015,baumgartnerSpatiallyTimeresolvedMagnetization2017}, with recent advancements extending static magnetic imaging to 5~nm~\cite{Battistelli2024, butcher2024,ranaThreedimensionalTopologicalMagnetic2023}. While this is sufficient to map typical DW profiles \cite{Battistelli2024}, the temporal resolution of soft-X-ray imaging with synchrotron sources is currently limited to tens of picoseconds \cite{mayr2024}. Cutting-edge transmission electron and scanning tunneling microscopy enable tracking and mapping of spin textures with a spatial precision down to 2~nm and 0.4~nm, respectively ~\cite{rubianodasilva2018, Moller2020,harvey2021,miyamachi2021, SPSTM,liu2025}, but access to femtosecond spin dynamics at such scales has not yet been reported. Time-resolved magnetic imaging with free-electron lasers offers femtosecond resolution with spatial resolution down to about 70~nm ~\cite{vonkorffschmisingImagingUltrafastDemagnetization2014}. Table-top high harmonic generation (HHG) sources feature exceptional time resolution and comprehensive polarization control in the XUV range \cite{kfirNanoscaleMagneticImaging2017,zaykoUltrafastHighharmonicNanoscopy2021,Zayko_2020}, which, nevertheless, presents challenges in reaching the spatial scales characteristic of domain walls. Specifically, common ferromagnetic materials such as Ni, Co and Fe feature absorption edges between 18 and 24~nm where magnetization-sensitive contrast can be obtained through X-ray magnetic circular dichroism (XMCD). Resolving DW properties at these wavelengths thus requires imaging with a resolution at or below the wavelength.

In this work, we track the local properties of nanoscale domain walls during ultrafast demagnetization using time-resolved imaging with a femtosecond XUV source, and reveal unexpected DW stability after intense laser irradiation. 
Reaching 13.5~nm sub-wavelength spatial and under 40~fs temporal resolution, we map spatial distributions of magnetic domain boundaries in ferrimagnetic TbCo and ferromagnetic Co/Pd thin films. Using detailed quantitative image analysis, we demonstrate spatial precision below 1~nm in determining DW widths and positions, and rule out transient DW broadenings or displacements for demagnetization levels up to approximately 60\%. The width of the spatially averaged domain wall profiles ($w_{avg}$) measured in each individual time frame as well as the average local DW width ($\bar{w}$) remains invariant within $\sim$ 1~nm across all time frames ($1\sigma <0.2$~nm).  At higher pump fluences inducing even stronger demagnetization levels, stochastic irreversible domain switching becomes prominent. Our results present a direct and spatially resolved picture of local spin dynamics during homogeneous laser-induced demagnetization, and establish upper bounds for reversible ultrafast domain wall dynamics caused by, for example, spin transport.

\begin{figure}
\centering
\includegraphics[width=\textwidth]{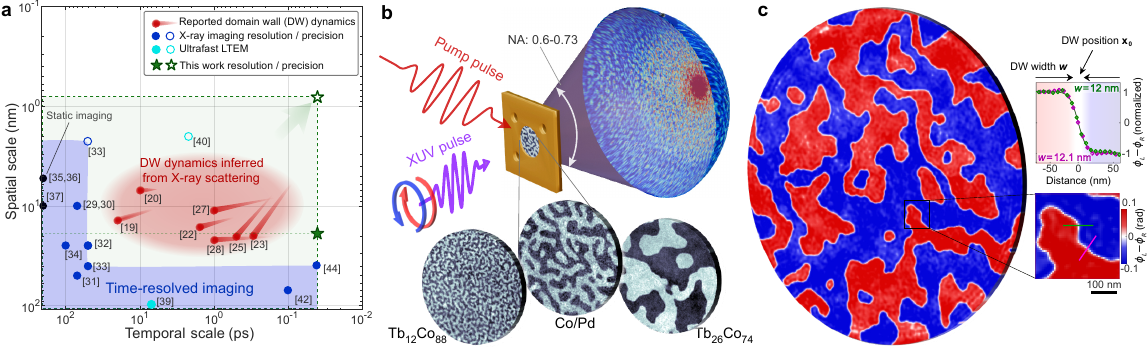}
\caption{Ultrafast sub-wavelength imaging of nanoscale spin textures using a table-top high-harmonic generation source.  \textbf{a}~Literature overview highlighting current magnetic imaging capabilities. Blue dots and circles represent time-resolved X-ray magnetic imaging, and the results obtained with Lorentz-transmission electron microscopy (LTEM) are shown in cyan. The estimated domain wall dynamics inferred from X-ray scattering experiments are indicated as red dots, with tails illustrating the reported temporal evolution. The spatiotemporal resolution and DW measurement precision achieved in this work are represented by green stars. \textbf{b}~Schematic of the experimental setup for optical-XUV pump-probe magnetic imaging. The sample, partially covered by an opaque mask defining a circular field of view, is uniformly excited from the magnetic film side by femtosecond laser pulses. Circularly polarized high-harmonic pulses capture the induced magnetization dynamics in real space via lensless imaging. \textbf{c}~Example static image of the magnetic domain pattern in Tb$_{26}$Co$_{74}$ alloy, recorded with 13.5~nm spatial resolution using a wavelength of 20.8~nm. The right panel shows a magnified region and two line profiles taken along the marked lines. Fitting with Eq.~(\ref{eq:Bloch-wall}) returns the local domain wall width parameter $w$ and position $x_{0}$ of the corresponding domain wall. }
\end{figure}

\section*{Results}
Our experiments employ an optical-pump, XUV-probe scheme, which is depicted in Fig.~1b and detailed in Methods. In brief, the sample is excited with femtosecond Ti:sapphire laser pulses and stroboscopically probed by time-delayed, circularly polarized XUV pulses at the Co $\text{M}_{2,3}$ edge (20.8~nm, $\sim$60 eV photon energy). An opaque mask with a 3-µm diameter circular field-of-view (FOV) is placed in close proximity to the magnetic film to confine the XUV illumination. The scattering signal is recorded in the far field, and the amplitude and phase of the electric field exiting the sample is reconstructed numerically using iterative phase retrieval. For each magnetic image, two sets of data are recorded with left- and right-circularly-polarized XUV pulses, respectively, to isolate the magnetic signal from non-magnetic contribution via XMCD. The illumination wavelength is tuned to maximize the phase difference contrast ($\phi_{L}(\mathbf{r}) - \phi_{R}(\mathbf{r})$) at the Co M$_{2,3}$ edge.
Example domain patterns in different materials and compositions are shown in Figs.~1b and 1c. Note that Fig.~1c presents an image of a ferrimagnetic sample close to its magnetic compensation point where opposing magnetic sub-lattices of Co and Tb cancel the net magnetization. Owing to the element-specific contrast of the imaging scheme, the domain texture is clearly resolved. A consistent isotropic resolution of 13.5~nm is achieved, with maximum scattering angles corresponding to a numerical aperture (NA) of 0.73. Further details on the quantification of the spatial resolution based on Fourier ring correlation, the phase-retrieval transfer function and a real-space analysis are provided in the Supplementary Information. 

\begin{figure}
\includegraphics[width=.7\textwidth]{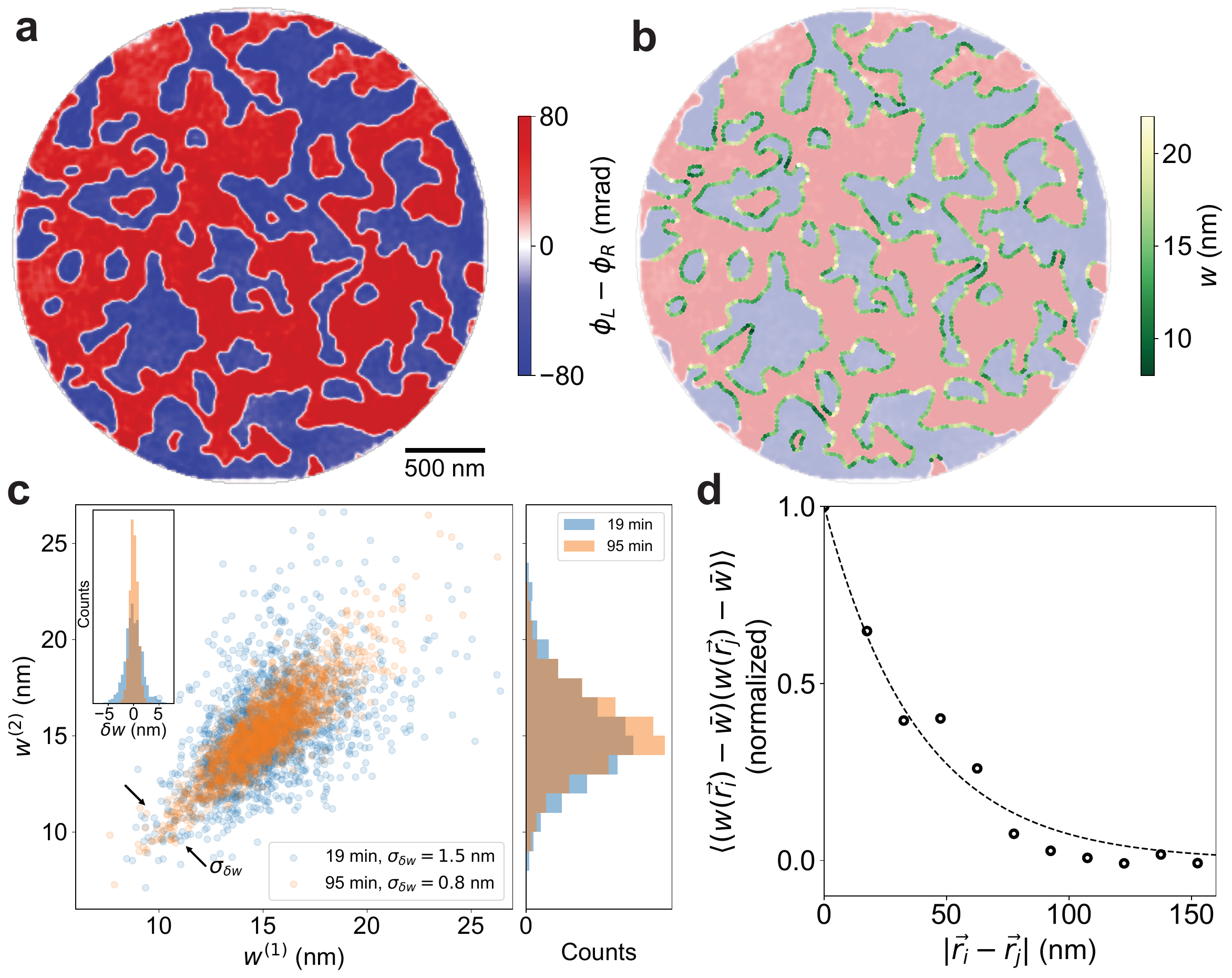}
\caption{Magnetic domain imaging before optical excitation (time delay $t<0$). \textbf{a} An example of a magnetic domain texture in a Tb$_{25}$Co$_{75}$ alloy sample. The local domain wall width across the field of view is shown in a green-yellow colormap in \textbf{b}. The fitted domain wall width is 15.2 $\pm$ 2.1~nm. \textbf{c} Correlation plots of the domain wall widths extracted from two independent measurements using a total of $\sim$19 (blue circles) and $\sim$95 minutes (orange circles) of exposure time per image. The spread of the off-diagonal distribution (inset) represents the difference in domain wall widths measured at the same locations in the two independent data sets, indicating the measurement precision (standard deviation $\sigma_{\delta w}$ reaching 0.8~nm). (d) Correlation function of DW width with respect to the distance between DW locations (circles), which is fitted with a single exponential function $\exp(-|\vec{r}_i-\vec{r}_j|/D_w)$ (dashed line) with decay length $D_w = 39$~nm.}
\end{figure}

Real-space images enable the direct extraction of local DW profiles as illustrated on the right side of Fig.~1c, where two example linecuts perpendicular to a domain wall are taken. For thin films with strong perpendicular magnetic anisotropy,
domain walls can be described by a
hyperbolic tangent function~\cite{hubertMagneticDomainsAnalysis1998,lemeshAccurateModelStripe2017,hennesLaserinducedUltrafastDemagnetization2020,zusinUltrafastPerturbationMagnetic2022}:
\begin{equation}
M_{z}(x)=M_{0}\mathrm{ tanh}\frac{x-x_{0}}{w},\label{eq:Bloch-wall}
\end{equation}
where $x$ denotes the position along a linecut perpendicular to the
DW contour, $M_{0}$ the out-of-plane magnetization of the
magnetic domain, $x_{0}$ the center position of the domain wall, and $w$
its width parameter. 
To obtain local DW profiles throughout the FOV, we first identify a coarse set of DW locations using contours with $\phi_L(\mathbf{r})-\phi_R(\mathbf{r})=0$. Coordinates for the profiles $\{x\}$ are thus lines perpendicular to the contours at every point.  To ensure unbiased measurements, the contours used to determine DW locations are identified first and kept fixed for DW width measurements in subsequent images.  At each location along the contour, the local width parameter $w$ and the center position $x_0$ are obtained by fitting Eq. (\ref{eq:Bloch-wall}) to the isolated DW profile.

Figures~2a and 2b show an example of domain-wall mapping in a Tb$_{25}$Co$_{75}$ sample. The local DW profiles are first identified across the FOV, and their extracted widths are shown in Fig.~2b using a green-yellow colormap. The measured DW width in this sample follows a normal distribution, with an average of $\bar{w}=15$~nm and a standard deviation $\sigma_w =2.1$ nm. It is important to note that, at a given spatial resolution, the precision of extracting local DW properties is limited by the signal-to-noise ratio (SNR) of the image, which is governed by the exposure time.
To determine the precision of the local DW width measurement, we compare the extracted DW widths of the same FOV (Fig.~2a) from two independently measured and separately reconstructed datasets. Figure~2c shows the correlation of the DW widths from the two datasets denoted as $w^{(1)}$ and $w^{(2)}$, respectively. Here, each dot represents the local DW width at the same location in the two datasets. The standard deviation of the histogram $\sigma_{\delta w}$ of the data points along the off-diagonal direction (see inset of Fig. 2c) indicates the degree of consistency between the two measurements. For the dichroic images acquired with a total of 1136~seconds of exposure time (blue circles), $\sigma_{\delta w}$ is 1.5~nm and it diminishes to 0.8~nm when the exposure time is increased by a factor of 5 (orange circles). Meanwhile, the distribution of DW width across the FOV (Fig.~2c, right panel) also narrows from 2.4~nm to 2.1~nm. As the standard deviation of the DW width distribution is larger than the measurement precision, and the decrease of $\sigma_{w}$ is insignificant compared to the increase in precision, the distribution of DW widths manifests an intrinsic variation of DW profiles across the sample. 

In Fig.~2b, characteristic changes in DW width along the contours occur on length scales substantially larger than the spatial resolution, suggesting contributions from local sample properties and domain topology. In addition, $w$ tends to shrink when there is another domain wall in close proximity. 
Similar correlations between DW width and domain size has been observed in 2D van der Waal magnets~\cite{birchHistorydependentDomainSkyrmion2022} and the variation in local DW properties was theoretically explored with the application of exchange bias~\cite{albisettiDomainWallEngineering2016}. For the TbCo sample here, the phenomenon can be explained by the inhomogeneity of atomic compositions, which leads to variations in the strength of exchange interaction across the sample. When exchange interaction is reduced, smaller domain sizes become more favorable, and the domain wall narrows \cite{hubertMagneticDomainsAnalysis1998}. To assess the inhomogeneity of DW widths, we plot the correlation function with respect to the distance between wall locations (Fig.~2d, black circles), which is defined as $\langle (w(\vec{r}_i) -\bar{w})(w(\vec{r}_j) -\bar{w})\rangle /\langle (w(\vec{r_i}) -\bar{w})^2\rangle$, with $\vec{r}_i$ denoting the \textit{i}\textsuperscript{th} DW location. The width correlation function can be modeled by a single exponential (dashed line) with decay length of 39~nm and thus can serve as a measure of sample inhomogeneity.

\begin{figure}
\includegraphics[width=\textwidth]{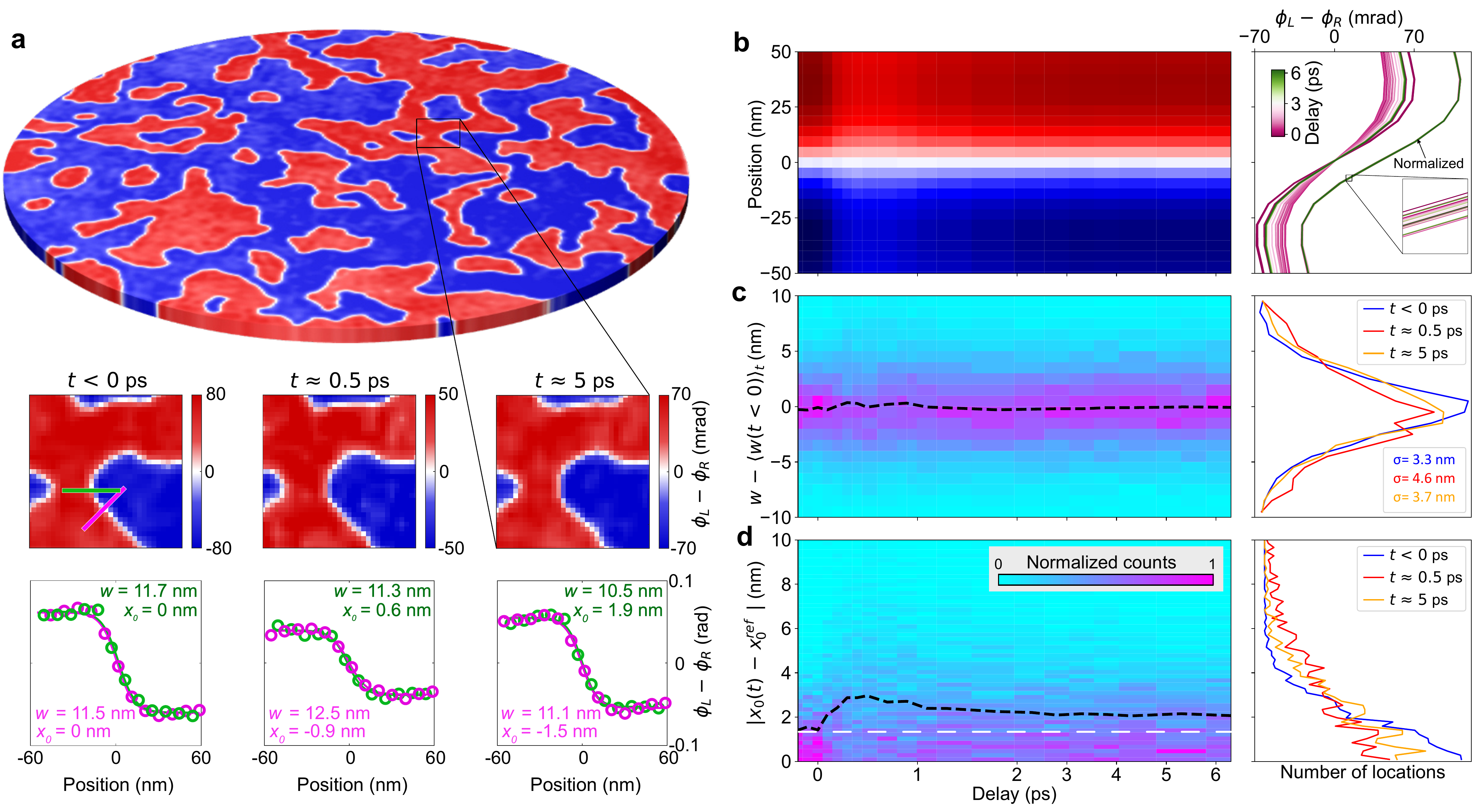}
\caption{Ultrafast imaging during demagnetization at a pump fluence of 6.2 mJ/cm$^2$, inducing $\sim 37\%$ demagnetization. 
\textbf{a} Magnetic domain pattern in Tb$_{25}$Co$_{75}$ sample before optical excitation ($t<0$). Magnified region shown in \textbf{a} highlights magnetic state before the optical excitation ($t<0$ ps), close to the maximum demagnetization ($t\approx 0.5$~ps), and during recovery ($t\approx 5$~ps), respectively. The lower panel shows the time evolution of an example domain wall during demagnetization. The locations of corresponding domain walls are shown with green and magenta lines on the image taken at a negative delay. The panel in \textbf{b} displays the averaged DW profiles across the FOV as a function of time delay. The right panel shows averaged DW profiles at different delay times. The curves on the right are the same curves normalized to the negative delay. \textbf{c}~Histograms of extracted width parameters $w(\vec{r},t)$ for all domain wall locations across the FOV compared to the corresponding width at negative delays ($\langle w(\vec{r},t<0)\rangle_t$). The panel on the right displays the three example histograms corresponding to the delays marked in \textbf{a}. \textbf{d} Time-dependent distribution of the magnitudes of displacements. The averaged magnitudes of displacements (black dashed line) exhibit a baseline of 1.5~nm (white dashed line), representing the precision in repeated measurements of individual domain walls.}
\end{figure}

Figure~3a illustrates the time-dependent magnetization dynamics of Tb$_{25}$Co$_{75}$ at three characteristic delays: the unpumped state ($t<0$ ps), near maximum magnetization suppression ($t\approx 0.5$ ps), and during relaxation ($t\approx 5$~ps). 
Throughout the FOV, we observe a decrease in overall magnetization within 0.4~ps, reaching $M_z(0.4 $~ps) $\approx 0.63 M_z(t<0$~ps), and a partial recovery after a few picoseconds. 
In this time-resolved dataset, we first analyze the temporal evolution of the spatially averaged DW profiles as shown in Fig.~3b and the resulting width $w_{avg}(t)$. The time-dependent profiles become indistinguishable from each other when normalized (Fig. 3b, right panel), and the statistical average of $w_{avg}(t)$ for positive and negative delays are both $15.1\pm 0.2$~nm, indicating that there is no statistically significant increase of the DW width after optical excitation. In addition, the standard deviation at negative times ($\sigma(w_{avg}(t<0))=0.2$~nm) reflects the high precision in measuring $w_{avg}$.

To systematically characterize potential changes in \textit{local} DW widths $w$, the histograms of the time-dependent difference $\Delta w = w(\vec{r},t) - \langle w(\vec{r}, t<0)\rangle_t$, representing the deviation of $w(t)$ from its unpumped value $w(t<0)$, are displayed in Fig.~3c. 
The corresponding time-dependent average of local changes in DW widths ($\langle \Delta w(\vec{r},t)\rangle_{\vec{r}}$) is plotted as a black dashed line (Fig.~3c, left panel). 
The distribution of $\langle \Delta w(\vec{r},t)\rangle_{\vec{r}}$ exhibits an average and standard deviation of $0.0\pm 0.2$ nm, consistent with the results from the averaged DW profiles (Fig.~3b). 
The right panel of Fig.~3c show the representative histograms of $\Delta w(t)$ at three different time delays. 
If DW broadening occurs at some locations, the time-dependent distribution of $\Delta w$ is expected to skew towards positive values for $t>0$ ps.
Compared to negative delays (blue line), the distributions at positive delays retain the symmetric shape with maxima centered around $\Delta w=0$, providing no indication of a systematic increase in DW width. The observed change in the distribution width in $\Delta w$ by $\sim 1.3$~nm as magnetization is suppressed (red line) is attributed to the diminishing XMCD contrast in the demagnetized state. This reduction in contrast effectively lowers the SNR of the magnetic image, leading to somewhat greater uncertainty in the extraction of DW properties (Fig. S4, Supplementary Information). 

Using an analogous approach to the width analysis, we now investigate potential DW displacements within the same dataset. The magnitude of displacements, $|x_{0}(t)-x_{0}^{\mathrm{ref}}|$, relative to their reference positions at negative delays, are computed for each location ($x_{0}^{\mathrm{ref}}=\langle x_0(t<0) \rangle_{t}$). Figure~3d displays their distributions across the FOV at different delays. The average displacement measured at $t<0$, is approximately 1.5~nm, representing the baseline uncertainty in determining DW positions (white dashed line). The average of absolute displacement as a function of time ($\langle |x_{0}(t)-x_{0}^{\mathrm{ref}}| \rangle$, black dashed line) increases to approximately 3~nm when the sample undergoes demagnetization, and subsequently decays as the magnetization recovers. 
The right panel of Fig.~3d displays the histograms of $|x_{0}(t)-x_{0}^{\mathrm{ref}}|$ at three representative delays. The distributions all peak at zero displacement and broaden at positive delays. However, this broadening can be attributed solely to the increased fitting uncertainty caused by reduced XMCD contrast. This assignment is supported by Monte Carlo simulations using artificial DW profiles (Supplementary Text S2 and Fig.~S5). Furthermore, binning experimental data from adjacent delays narrows the width of the distributions and lowers the magnitude of the displacements by about 50\% (Figs.~S6, Supplementary Information). 
Therefore, the averaged magnitude of DW displacement must be less than 3~nm and is likely to be considerably smaller.

\begin{figure}
\includegraphics[width=0.7\paperwidth]{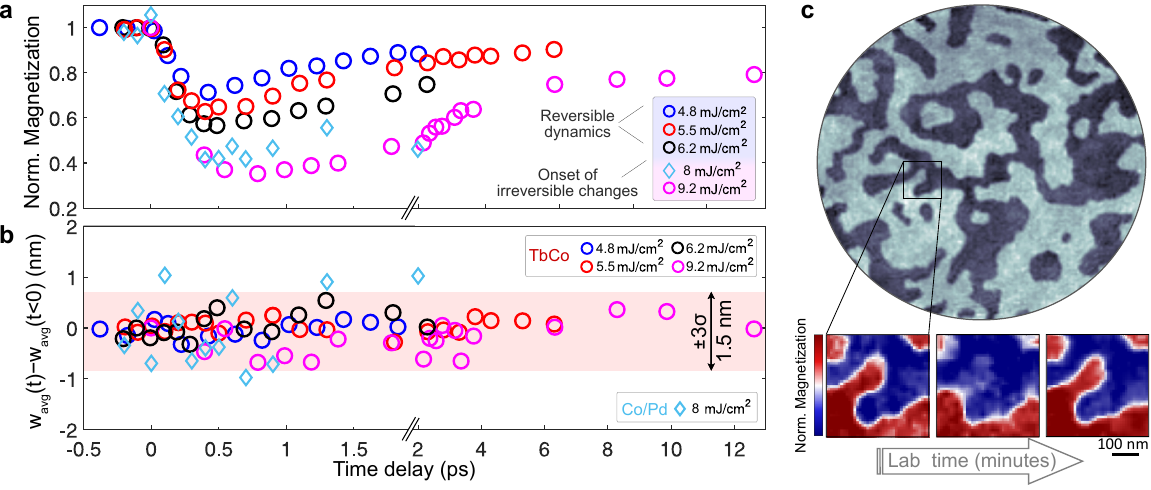}
\caption {Domain wall stability at various pump fluences. \textbf{a} Spatially averaged magnetization ($|M_z|$) across the FOV, normalized to the values at negative delays. \textbf{b}  Time-dependent changes in widths of the spatially averaged DW profiles compared to negative delays ($w_{avg}(t)-\langle w_{avg}(t<0)\rangle_t$) for each pump fluence shown in \textbf{a}. The combined distribution of $w_{avg}(t)-\langle w_{avg}(t<0)\rangle_t$ for all measured fluences with TbCo exhibits a standard deviation $\sigma=0.25$~nm, and the corresponding $\sigma$ for Co/Pd is 0.7~nm. The red shaded area indicates the $\pm 3\sigma$ interval of the combined data set recorded with TbCo sample. \textbf{c} An example of 
delay-independent, local stochastic domain switching in Tb$_{25}$Co$_{75}$ at a pump fluence of 9.2~mJ/cm$^2$. The three images were acquired consecutively, with exposure time over several minutes each, i.e. $>10^5$ pump-probe laser shots.}
\end{figure}

With the established precision in resolving DW properties, we further investigate potential DW dynamics at different pump fluences in the Tb$_{25}$Co$_{75}$ sample, focusing on $w_{avg}$. 
The time-dependent magnetization and change of $w_{avg}$ with respect to its unpumped value ($\langle w_{avg} (t<0)\rangle_t$) under different pump fluences are displayed in Figs.~4a and 4b, respectively.
Notably, $w_{avg}(t)$ remains largely unchanged irrespective of fluence.
For Tb$_{25}$Co$_{75}$, the distribution of $w_{avg}(t)-\langle w_{avg} (t<0)\rangle_t$ presents an average and standard deviation of $0.0\pm 0.3$~nm under pump fluences up to 9.2~mJ/cm$^2$. 
The Co/Pd sample exhibits similar behavior, though with a larger standard deviation of 0.7~nm due to its lower XMCD contrast. Since both DW broadening and displacement would result in an increase in $w_{avg}$, its unchanging width suggests that the overall domain structures remain stable for magnetization suppression down to approximately 40\% ($M_z\approx 0.4 M_z(t<0)$, Figs.~4a and 4b).

At high pump fluences inducing demagnetization above approximately 50\%, irreversible changes in the magnetic domain patterns emerge in certain regions of the samples. Figure~4c illustrates the onset of these changes at a fluence of $\sim 9$~mJ/cm$^2$ in Tb$_{25}$Co$_{75}$. In the magnified region, magnetization switching is observed. The frequency, area, and number of locations exhibiting such behavior, increase for larger pump fluences. 
Note that for the aforementioned results of DW widths and displacements (Figs.~4a and 4b), regions with irreversible changes are first identified and omitted during data processing. 

\section*{Discussion}
The robustness of magnetic domain walls and the occurrence of stochastic irreversible switching under strong excitation provide a valuable perspective for existing reports on DW dynamics. Many of the X-ray and XUV scattering studies suggesting ultrafast spatiotemporal dynamics of domain walls were conducted either at very high pump fluences or revealed a strong fluence dependence, occasionally exhibiting threshold-like behavior when several fluences were explored~\cite{pfauUltrafastOpticalDemagnetization2012,zusinUltrafastPerturbationMagnetic2022,jangidExtremeDomainWall2023}. 
By contrast, experiments carried out at moderate to low pump fluences with $\leq 20$\% demagnetization showed no scattering peak variations indicative of DW dynamics in the first few picoseconds~\cite{hennesLaserinducedUltrafastDemagnetization2020,weder2020}. 
Hence, the observed DW stability in this work is consistent with the findings of X-ray scattering data at low fluence. 
Moreover, this comparison suggests that previously observed changes in scattering angles and variations in the diffraction ring intensity from magnetic domain patterns may involve or coincide with irreversible changes in the domain structure, particularly in regions where the domain arrangement becomes highly unstable. Our data at high pump fluences also indicate that regions consisting of small domains are more prone to local switching, which may result in intensity suppression at high diffraction angles and be subsequently interpreted as shrinkage of diffraction rings in scattering experiments.
Such transient instability in magnetic domains could serve as a precursor for full magnetic switching or indicate the existence of a transient state in which local magnetic moments rapidly fluctuate before stabilizing into a new configuration. 
Similar behavior has been proposed in the extreme case of a laser-induced ferromagnetic-to-paramagnetic phase transition~\cite{suturinShortNanometerRange2023}. 
These fluctuations may outline a potential pathway to ultrafast switching, where domain wall rigidity could play a critical role in the onset of switching, mediating the transition between magnetic configurations and potentially facilitating stable, ultra-small and ultrafast magnetic toggle \cite{steinbach2024}. 

In conclusion, the presented results provide narrow upper bounds for the broadening and displacement of domain walls under reversible ultrafast demagnetization induced by homogeneous optical excitation. Our observations imply a limited influence of spin transport and are consistent with the reported propagation lengths in typical spintronic emitters \cite{seifertSpintronicSourcesUltrashort2022}. While our measurements are not sensitive to out-of-plane spin currents, they place stringent constraints on lateral spin-polarized carriers and, consequently, on isotropic carrier dynamics. Our findings on domain wall width is consistent with the estimation from quasi-thermal heating (see Supplementary Text S3) and align with the previous reports on in-plane spin dynamics with localized excitations \cite{weder2020}.

The demonstrated sub-wavelength, full-field XUV microscopy provides a powerful platform for exploring ultrafast magnetism at its fundamental length and time scales. The element specificity enables access to complex systems even near the magnetic compensation point, as shown for TbCo, and facilitates studying compensated and altermagnetic materials  \cite{Hariki2024}. Beyond magnetism, the method offers real-space insights into nanoscale chemical and structural dynamics \cite{stinsonImagingNanoscalePhase2018,johnson2022,Heinrich2023, Domroese2024}, and even holds promise for high-resolution metrology in the context of extreme ultraviolet lithography. This unique combination of spatiotemporal precision, material sensitivity, and broad applicability renders XUV microscopy a highly promising tool for future ultrafast materials science.

\section*{Methods}
\paragraph*{Sample preparation.}
The ferromagnetic sample used is a Co/Pd multilayer with the following structure: Co(0.5 nm)/Pd(0.75 nm) repeated 9 times. The ferrimagnetic films are Tb$_{1-x}$Co$_x$ alloys with a thickness of 20~nm. For all datasets except Fig.~1b, $x$ = 0.74 or 0.75, and for Fig.~1b, $x$ is varied between 0.74 and 0.88. 
All samples exhibit strong perpendicular magnetic anisotropy with easy-axis magnetization pointing out-of-plane. 
The samples were prepared at room temperature by DC magnetron sputtering from elemental targets onto 200-nm thick silicon membranes. 
The sputtering process was carried out with argon gas at a working pressure of $3.5\times 10^{-3}$~mbar in a vacuum chamber (base pressure $<10^{-8}$~mbar). 
To protect the films from oxidation, 2~nm of Pd or 5~nm of $\text{Si}_{3}\text{N}_{4}$ was used as a cover layer. 
The layer thicknesses were estimated using areal densities measured by a calibrated quartz balance before the deposition.

For the imaging mask, 200~nm of gold was deposited on the backside of the samples via thermal evaporation. At locations separated by $>300\, \mu$m, the gold film was selectively etched with a focused ion beam (FIB) to create circular apertures with a 3~µm diameter as the field of view (FOV). Several auxiliary holes, with diameters ranging from 200 to 600~nm, were milled through the entire structure to facilitate scattering data collection at high numerical aperture. On some masks, two additional reference holes (with diameters of 200 and 300~nm) were fabricated to serve as reference sources for Fourier transform holography. The holographic reconstruction is used only to provide the initial guess for the real-space support in the phase retrieval process. Compared to the mask design in Ref.~\cite{zaykoUltrafastHighharmonicNanoscopy2021}, the arrangement of auxiliary holes is optimized to meet increased longitudinal coherence requirements by reducing the total size along the axis of spectral dispersion (see Fig.~S7). 

\paragraph*{Experimental setup.}
The system encompasses a commercial Ti:sapphire laser with wavelength centered at around 800~nm, a repetition rate of 1~kHz, and pulse duration of 33~fs. The output of the laser is split into a pump and a probe arm. The pump beam, with pulse energy of $\sim$100 $\mu$J, is time-delayed using a delay stage. Its intensity is tuned with a neutral density filter for fluence-dependent measurements. The probe arm, which drives the high-harmonic generation (HHG) process, has a pulse energy of 1.6~mJ. A $\beta$-barium borate (BBO) crystal is inserted into the probe beam path to generate 400 nm light that co-propagates with the 800~nm pulses. These pulses are delay-compensated using a pair of calcite plates and converted to circular polarization by an achromatic quarter-wave plate. The counter-rotating bi-color, circularly-polarized pulses then enter a He-filled gas cell, where circularly polarized XUV pulses are produced via HHG~\cite{Kfir2016}. The dynamical symmetry of the HHG process leads to every third harmonic order being suppressed. By fine-tuning the time-delay between the pulses at attosecond time scales, different suppression rules can be selected as described in Ref. \cite{Zayko_2020}. 
After the gas cell, a 200-nm thick Al foil isolates the XUV light from the fundamental bi-color beam. The XUV beam is wavelength-dispersed and refocused onto the sample using an XUV monochromator. To excite the Co M$_{2,3}$ edge, a wavelength of approximately 20.8~nm is selected by a slit placed in front of the sample. The XUV light diffracted by the sample is collected by a CCD camera, with a 400-nm thick Al filter placed in front to block stray light from the pump beam. The helicity of the XUV beam is controlled by adjusting the rotation angle of the quarter-wave plate.

To optimize the XUV beam for imaging, we first fine-tune the HHG parameters for optimal flux. Secondly, wavefront optimization is done by adjusting the XUV monochromator in all six degrees of freedom (3 translational and 3 rotational axes) to minimize higher-order aberrations and maximize the XUV intensity on the sample. Finally, we use a double pinhole to optimize the coherence properties of the XUV beam \cite{changCoherenceTechniquesExtreme2002}.

The laser pump fluence is calculated by first measuring the focal spot size at the sample position using a pinhole with 15~$\mu$m diameter. By translating the sample film and measuring the transmission of the pump beam through the pinhole, the full-width-at-half-maximum of the pump beam at vertical ($w_a$) and horizontal direction ($w_b$) are extracted by fitting the transmission profile with a Gaussian function. The pump fluence can thus be described as $F=(4E\ln 2)/(\pi w_a w_b),$ where $E$ is the pulse energy.

\paragraph*{Data collection and iterative phase retrieval.} 
The diffraction data are collected using $1024\times 1024$ pixels for time-resolved data and $2048\times 2048$ for static high-resolution images (Figs.~1b,c). Dark images without XUV exposure were taken and subtracted from the diffraction image. To account for the high dynamic range of the scattering signal, several acquisitions with varying exposure times are recorded and merged. Short exposure times capture only low-frequency components, typically ranging from 0.5 to 5 seconds, depending on the NA used, while longer exposure times are typically between 20 and 60 seconds. For example, to obtain a single real-space image of the time series shown in Fig.~3a, the average of 8 acquisitions at 0.5 seconds each was combined with an averaged image of 10 acquisitions at 28 seconds each to form a single high-dynamic range diffraction pattern. This approach is sufficient to achieve high-quality real-space image with sub-wavelength resolution for a single helicity. Nevertheless, the magnetic signal in such an image is weak and overwhelmed by non-magnetic contributions. To isolate the magnetic signal, an identical dataset with opposite circular polarization is acquired. Therefore, the total exposure time required for a single time-delay image in Fig.~3 is  $(0.5\, \mathrm{s} \times 8+28\,\mathrm{s} \times 10) \times 2 = 568$ seconds. Longer exposure times improve the image quality and thus the precision of DW width and position. Before conducting iterative phase retrieval, the diffraction data are projected onto the Ewald sphere to account for curvature at high scattering angles \cite{rainesThreedimensionalStructureDetermination2010}.

The autocorrelation from the Fourier transform of the diffraction image is used to create the imaging support for iterative phase retrieval. Due to the presence of distant reference holes (Fig. ~6), the autocorrelation contains cross-correlation terms between the mask and the reference holes. This cross-correlation (the Fourier transform holography reconstruction of the object) provides the exact shape of the mask, with resolution governed by the size of the smallest reference hole. The real-space pixel size of this reconstruction is inversely related to the spatial frequency at the edge of the detector $q$, given by $\text{pixel size} = 1/(2q)$. Since the distance between the reference holes is known from the FIB design, the spatial frequency at the edge of the detector can be obtained without exact knowledge of the sample-to-detector distance. 

The iterative phase retrieval process begins with data recorded under left circularly polarized light and a random initial guess. First, we run 100 iterations of the RAAR algorithm \cite{lukeRelaxedAveragedAlternating2005}, with a relaxation parameter of 0.9. In the following 300 iterations, the relaxation parameter is gradually reduced from 0.9 to 0.5. An additional 20 iterations of error reduction (ER) \cite{fienupReconstructionObjectModulus1978} are then applied to finalize preliminary reconstruction of left circularly polarized light. 

After this initial run, the real-space support is updated via thresholding the real-space image. The entire process is then repeated three more times to improve the accuracy of the real-space support. The final support is subsequently used for reconstructing all data recorded with this mask and numerical aperture, e.g. all frames within one time-resolved experiment. 

The preliminary reconstruction process assumes monochromatic, perfectly coherent light and accounting for partial coherence due to imperfections in the optical system is required to lower the reconstruction error. To achieve this, an additional 200 RAAR iterations are performed, with the mutual coherence function estimated every 10th iteration using 15 steps of Richardson-Lucy deconvolution~\cite{clarkHighresolutionThreedimensionalPartially2012,Battistelli2024}. 
The final reconstruction then serves as the initial guess for reconstructing data recorded with the opposite helicity, for which only 20 iterations of ER are required.

The phase difference of the reconstructions recorded with opposite helicities forms the phase-contrast image ($\phi_L-\phi_R$) as shown in Figs.~1-4, which is proportional to the out-of-plane magnetization of the sample \cite{kfirNanoscaleMagneticImaging2017}. Although the first image can serve as an initial guess for subsequent reconstructions, e.g. in a time-dependent measurement, such practice is avoided to prevent potential bias towards a close or similar solution. In this work, all time-resolved data sets are reconstructed independently using random initial guesses, with no overlap between data recorded at different times. 

It is important to note that the real-space support determines the exact plane of the reconstructed exit surface wave. When the support is loose, the final reconstruction may appear blurred or defocused and must be numerically propagated to the correct plane at the exit surface.  With a well-defined, tight support, the reconstruction algorithm reliably converges to an effectively identical solution across the momentum space, regardless of the initial guess. The convergence is verified by the phase retrieval transfer function and Fourier shell correlation analysis in the Supplementary Fig.~S1, where both metrics exceed the threshold criteria across the entire momentum space, indicating isontropic diffraction limited resolution down to 13.5~nm.

\paragraph*{Fitting of domain wall profiles.} Before the extraction DW properties, a super-Gaussian filter is applied to smooth the edges of the far-field data (see Supplementary Information). The resulting magnetic image is then interpolated and mapped onto a grid with spacing equals to half the real-space pixel size. The positions of domain walls are localized using contours with $\phi_L(\mathbf{r})-\phi_R(\mathbf{r})=0$, with points on the contours separated by the grid size. The direction of each linecut perpendicular to the domain wall is determined by taking the gradient of the image ($\nabla(\phi_L(\mathbf{r})-\phi_R(\mathbf{r}))$). Before taking the gradient, the image is first denoised using a Gaussian filter $\exp((|\mathbf{r}|^2/(2\sigma^2)))$ with $\sigma=1.5$ grid size. Each domain wall profile along the linecut $M_z(x)$ is obtained using 2D linear interpolation of the image with $x$ ranging between -7 and 7 grid pixels. To further reduce the effect of noise on the fitting, we bin domain wall profiles at every 3 neighboring domain wall positions assuming that domain wall properties do not change drastically within $\sim$28~nm distance along the DW.
The fitting is conducted with nonlinear least squares method. For small magnetic domains with sizes close to the length of linecut $x$, the wall on the opposite side of the neighboring domain may affect the magnitude of $M_z(x)$ at the ends of the linecuts. Therefore, the fitting is weighted such that points with $x\in[-2,2]$ pixel ($x\in[-4,4]$ grid size) are weighted 8 times compared to other points on the linecut.

\begin{acknowledgments}
This work is funded by the Deutsche Forschungsgemeinschaft (DFG, German Research Foundation) through the Gottfried Wilhelm Leibniz program. H.-T. C. acknowledges support from the Alexander von Humboldt Foundation.
\end{acknowledgments}

\bibliographystyle{apsrev4-2}
\bibliography{references}
\end{document}


\title{Supplementary Information to ``Revealing domain wall stability during ultrafast demagnetization''}
\author{Hung-Tzu Chang}
\thanks{H. -T. C. and S. Z. contributed equally to this work.}
\affiliation{Max Planck Institute for Multidisciplinary Sciences, 37077 G\"ottingen,
Germany}
\author{Sergey Zayko}
\thanks{H. -T. C. and S. Z. contributed equally to this work.}
\email{szayko@gwdg.de}
\affiliation{Max Planck Institute for Multidisciplinary Sciences, 37077 G\"ottingen,
Germany}
\author{Timo Schmidt}
\affiliation{Institute of Physics, University of Augsburg, 86135 Augsburg, Germany }
\author{Ofer Kfir}
\affiliation{School of Electrical Engineering, Tel Aviv University, Tel Aviv 69978, Israel}
\author{Murat Sivis}
\affiliation{Max Planck Institute for Multidisciplinary Sciences, 37077 G\"ottingen,
Germany}
\affiliation{4\textsuperscript{th} Physical Institute, University of G\"ottingen,
37077 G\"ottingen, Germany}
\author{Johan H. Mentink}
\affiliation{Radboud University, Institute for Molecules and Materials, Nijmegen, 6525 AJ, Netherlands}
\author{Manfred Albrecht}
\affiliation{Institute of Physics, University of Augsburg, 86135 Augsburg, Germany }
\author{Claus Ropers}
\email{claus.ropers@mpinat.mpg.de}
\affiliation{Max Planck Institute for Multidisciplinary Sciences, 37077 G\"ottingen,
Germany}
\affiliation{4\textsuperscript{th} Physical Institute, University of G\"ottingen,
37077 G\"ottingen, Germany}
\maketitle

\section{Characterization of spatial resolution}
In coherent diffractive imaging, the real-space pixel size is determined by the numerical aperture and wavelength. However, the actual spatial resolution does not necessarily reach this limit. Instead, it is governed by the maximum scattering angle that can be reliably reconstructed, which is  the highest diffraction signal recorded with sufficient signal-to-noise ratio (SNR).

After real-space reconstruction, image quality and resolution can be assessed using far-field methods such as the phase retrieval transfer function (PRTF) and Fourier shell correlation (FSC or Fourier ring correlation for 2D data), which evaluate phase consistency and compare correlations from independent measurements. If far-field phases are consistently reconstructed across the entire momentum space—i.e., for all recorded spatial frequencies—the image can be considered diffraction-limited, with the true pixel resolution defined by the momentum transfer at the edge of the detector \cite{chapman2006}. 

A similar approach applies to FSC analysis, where the correlation coefficient is compared to threshold criteria as a function of spatial frequency. The two most commonly used thresholds are the 0.143 and half-bit criteria \cite{rosenthalOptimalDeterminationParticle2003,vanheelFourierShellCorrelation2005}. However, far-field estimates should be validated with real-space analysis, such as measuring the smallest resolvable feature or using the knife-edge technique.

\begin{figure}
\includegraphics[width=\columnwidth]{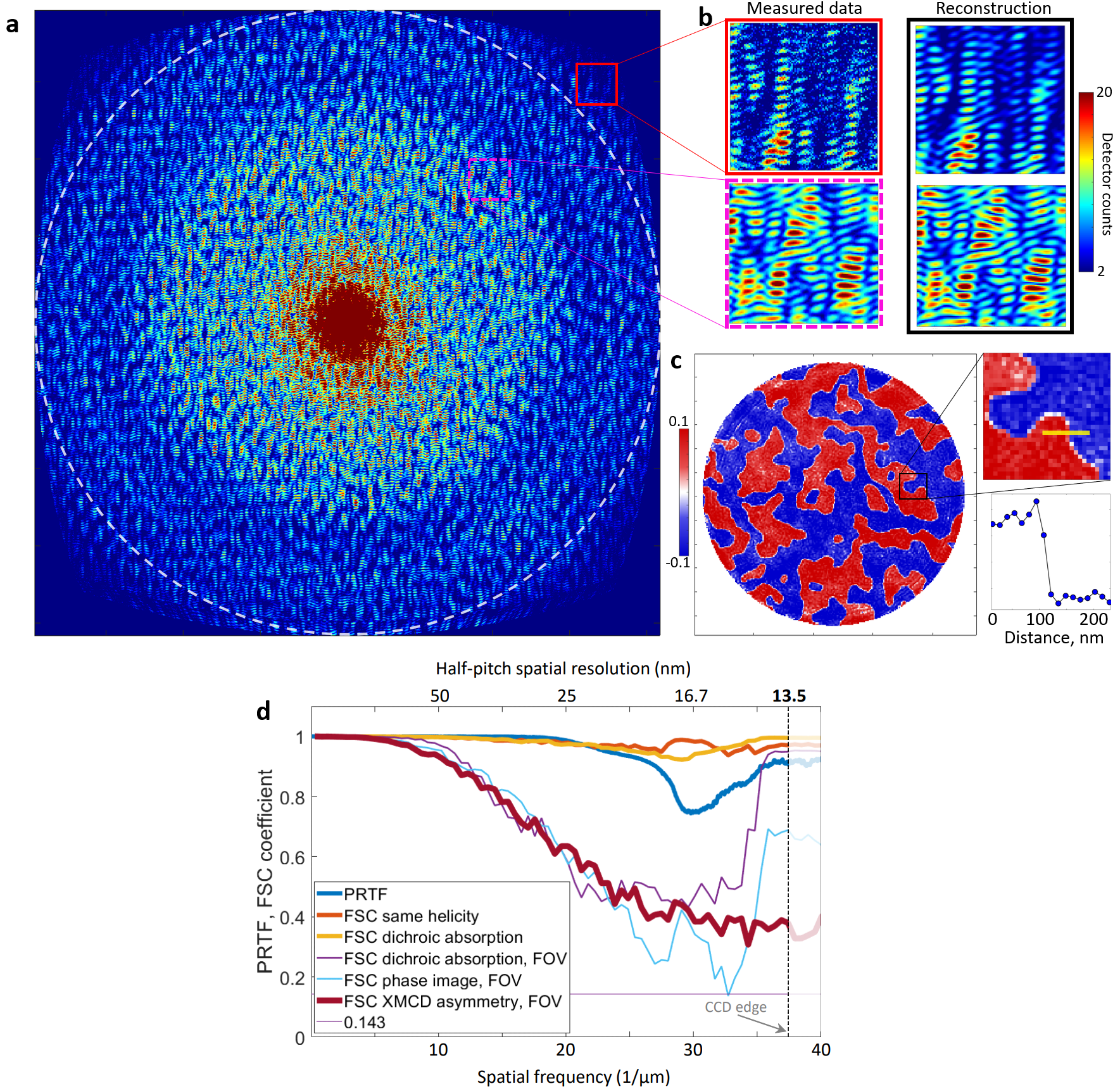}
\caption {Spatial resolution verification. \textbf{a} Far-field scattering data recorded with a 10-minute exposure time. The dashed line indicates the spatial frequency at the edge of the CCD (37$~\mu\mathrm{m}^{-1}$, defining a real-space pixel size of 13.5~nm. \textbf{b} Comparison of the measured data with the reconstructed far-field shows good agreement, even in the corners of the detector, where the SNR is weakest. \textbf{c} Corresponding reconstruction of the isolated magnetic signal shown as reconstructed in true-pixel without interpolation. The zoom-in  regions highlights where DWs are resolved within a single pixel transition, indicating true-pixel resolution in real-space. \textbf{d} Far-field analysis using the phase retrieval transfer function (PRTF) and Fourier shell correlations (FSC) demonstrates reliable imaging across all recorded spatial frequencies and different types of image contrast.}
\label{fig:resolution}
\end{figure}

Figure \ref{fig:resolution}a shows a typical diffraction pattern recorded with this setup and the mask design (see Fig. \ref{fig:mask}), optimized for the uniform filling of momentum space. The white dashed line highlights the spatial frequency at the edge of the detector, $q= \frac{1}{2\times (\mathrm{pixel\, size})} \approx 37~\mu\mathrm{m}^{-1}$, which determines the real-space pixel size of 13.5~nm. Figure \ref{fig:resolution}b compares the scattered data with the reconstructed signal, showing good agreement between the measured and reconstructed data. Even at spatial frequencies beyond $q = 37~\mu\mathrm{m}^{-1}$, where the scattering signal is extremely weak, the underlying diffraction fringes are successfully recovered (cf. Fig. \ref{fig:resolution}b). For spatial frequencies below $37~\mu\mathrm{m}^{-1}$, the retrieved data are indistinguishable from the measurement. Figure~\ref{fig:resolution}c shows the magnetic image from the far-field data in Fig.~\ref{fig:resolution}a as reconstructed without interpolation.  Here, the magnetic signal within the circular field-of-view is shown as the phase difference between two data sets recorded with opposite helicities. The magnified region highlights multiple domain wall (DW) profiles, which are resolved within a single pixel transition, demonstrating isotropic true-pixel resolution in real space. Far-field analysis is shown in Fig.~\ref{fig:resolution}d, where the PRTF value computed for the single helicity image (Fig. \ref{fig:resolution}a) remains above 0.75 across the momentum space. This confirms that all spatial frequencies are consistently reconstructed irrespective of the initial guess. Additionally, several FSC plots, computed from different types of image contrast for (i) the entire exit-surface field including the signal from auxiliary holes, and (ii) isolated circular FOV images, also support a diffraction-limited resolution of 13.5~nm irrespective of the image contrast mechanism used.

\section{The accuracy and precision of extracted domain wall properties} \label{sec:accuracy}
\begin{figure}
\includegraphics[width=\columnwidth]{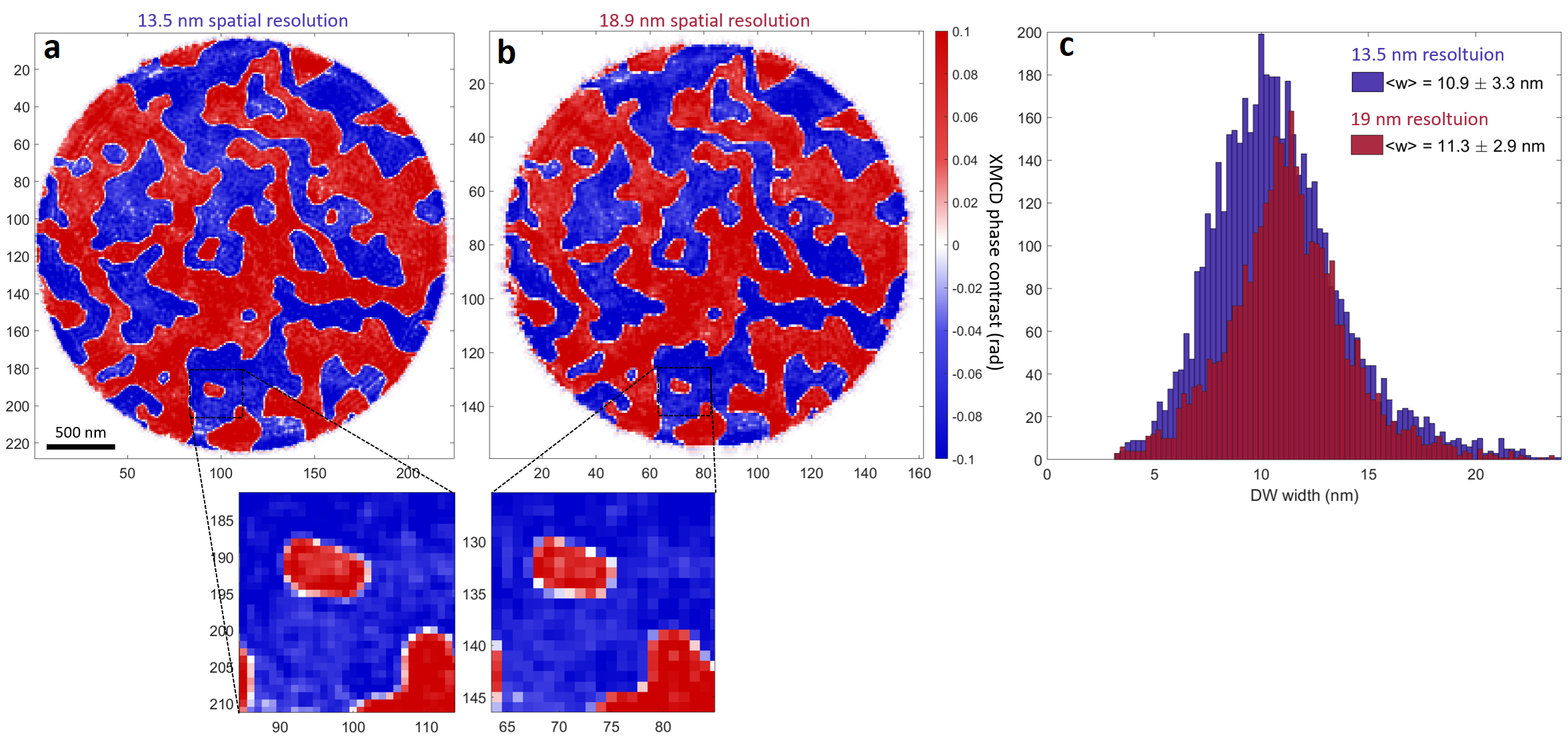}
\caption {Comparison of domain wall measurement analysis for the data recorded at relaxed imaging resolution. \textbf{a} Magnetic state recorded with 13.5~nm spatial resolution using XMCD phase contrast imaging. \textbf{b} Same sample imaged with 18.9~nm spatial resolution. \textbf{c} Local DW width analysis results for both images. The relative change in resolution is substantially larger than the relative change in the measured mean values $\bar{w}$, which remains within the uncertainty of the absolute DW width between independent datasets recorded at different NA (see text). The relative change in the resolution used is substantially larger than the relative change of the measured mean values $\bar{w}$ which is within the uncertainty of the absolute value of DW width between independent sets (see text).}
\label{fig:differentNA}
\end{figure}

\paragraph*{Results with different numerical apertures.} To characterize the accuracy of the measured DW width and its potential overestimation due to the inherent point spread function of the imaging system, we performed a set of measurements on the same sample using a reduced numerical aperture (NA) that limits spatial resolution to 18.9~nm (cf. Fig.~\ref{fig:differentNA}, note that this is the same sample shown in Fig.~\ref{fig:resolution}, now shown in inverted colormap). The average DW width measured in several thousand linecuts across the FOV is 10.9~nm, with a standard deviation of 3.3~nm when the sample is imaged with 13.5~nm spatial resolution. For the dataset recorded with reduced NA, the mean value $\bar{w}$ increases to 11.3~nm, which is significantly lower than the difference in resolution. Since higher resolution imaging requires substantially longer exposure time to ensure sufficient SNR for high spatial frequencies, all time-resolved data shown in, Figs.~2-4 of the main text were recorded at the reduced spatial resolution of 18.9~nm, while still maintaining sufficient precision for local domain wall width measurements.

\paragraph*{Effect of noise.} The precision of the extracted DW width and position is strongly influenced by the SNR of the image. High-frequency noise (see the typical linecut in Fig.~\ref{fig:resolution}c) primarily originates from the noise in the diffraction data at the highest scattering angles (cf. Fig.~\ref{fig:resolution}b). This can be partially suppressed by applying a super-Gaussian filter to the edges of the far-field data before real-space analysis. The filter consists of a circular, quasi-binary mask with diameter matching the size of the detector. The value of the mask inside the circle is 1 and outside 0.3. To de-noise the signal, the mask is first smoothed by a super-Gaussian filter 
$$G(\mathbf{k})=\left( e^{-\frac{k^2}{2\sigma_g^2}} \right)^P,$$
with $\sigma_g=15$ pixels and power $P=3$. The final image is obtained by multiplying the reciprocal-space image of the final reconstruction with the filter and performing an inverse Fourier transform. 
The application of the filter reduces $\sigma_{\delta w}$ (see Fig. 2c in the main text) by up to 2~nm while increasing the mean value of DW widths $\bar{w}$ by 2–4\%. As we are mainly interested in the \textit{relative} changes of DW properties in time-dependent measurements, such slight increase in overall DW width does not affect our conclusions. 

\begin{figure}
\includegraphics[width=.8\textwidth]{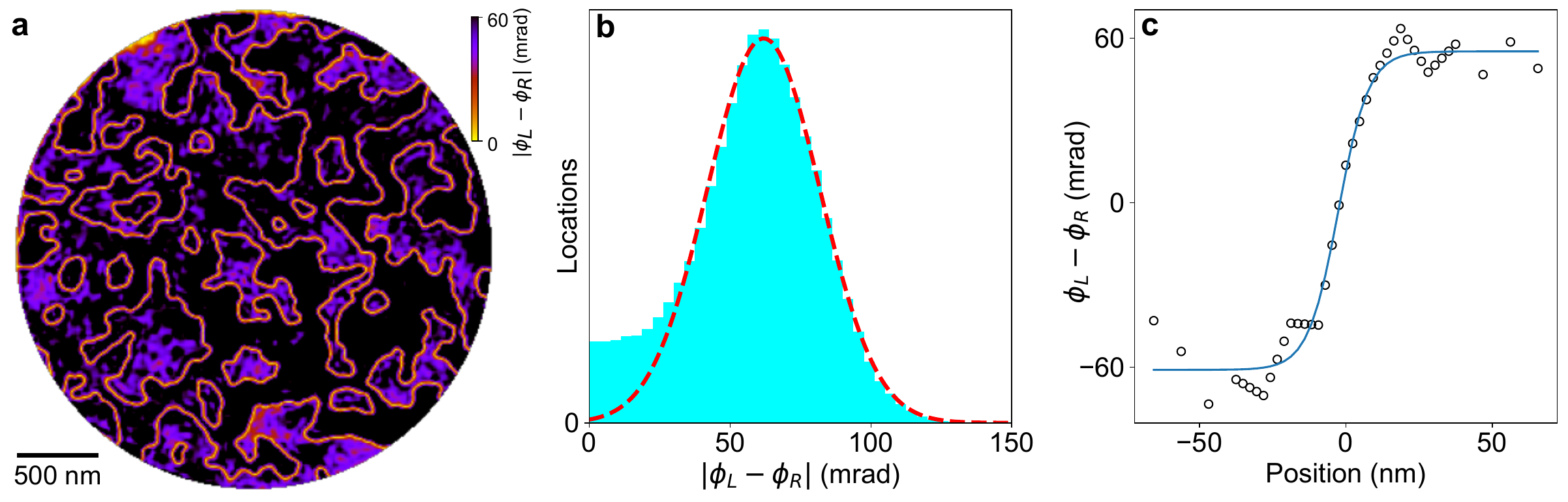}
\caption{Extraction of noise level from acquired images for Monte-Carlo simulations. \textbf{a} Absolute value of phase contrast of a reconstructed magnetic image in unpumped state. \textbf{b} Extraction of noise level by fitting a Gaussian (red dashed line) to the histogram of absolute phase contrast in cyan. \textbf{c} An example of simulated DW profile (black circles) and its fit with Eq. (1) in main text (blue line).}
\label{fig:MCscheme}
\end{figure}

To further examine the impact of noise on extracted DW properties, we perform Monte-Carlo calculations with simulated DW profiles. First, we extract the noise level of the measured real-space images by fitting a Gaussian to the histogram of absolute magnetic contrast $|\phi_L-\phi_R|$ (Figs.~\ref{fig:MCscheme}a,b). Second, a DW profile is generated with Eq. (1) in the main text with artificially set $w$ and $M_0$ according to fitting results from unpumped data. Here, the position $x_0$ is set to 0. Third, Gaussian noise with the noise level from the first step is added to the profile, and finally, we perform a fitting with Eq.~(1) in the main text to see the deviation of fitting parameters from the original value (Fig. \ref{fig:MCscheme}c). To account for partially demagnetized conditions, $M_0$ is modified to reflect the level of demagnetization. The width $w$ for the generated profiles are set to 15~nm, in accordance with empirical value. For each $M_0$, we generate 10000 DW profiles with the aforementioned procedure and obtain the distribution of fitted width ($w_\mathrm{MC}$) and position ($x_{0,\mathrm{MC}}$). Note that the result of the Gaussian fitting to the histogram of $|\phi_L-\phi_R|$ contains contribution from both experimental noise and intrinsic variations in magnetization due to sample inhomogeneity. Such potential overestimation of noise level will cause the standard deviations of the distributions of $w_\mathrm{MC}$ and $x_{0,\mathrm{MC}}$ to be larger than those from experimental data. Figure \ref{fig:displacementMC} shows a comparison of the empirical distributions of $|x_0-x_{0}^{\mathrm{ref}}|$ (cf. Fig.~3d in main text) and the results from Monte-Carlo calculations with 10000 samples. The good agreement between experimental results and simulation confirms that the distribution of $|x_0-x_{0}^{\mathrm{ref}}|$ is mainly contributed by noise.
\begin{figure}
    \centering
    \includegraphics[width=0.4\linewidth]{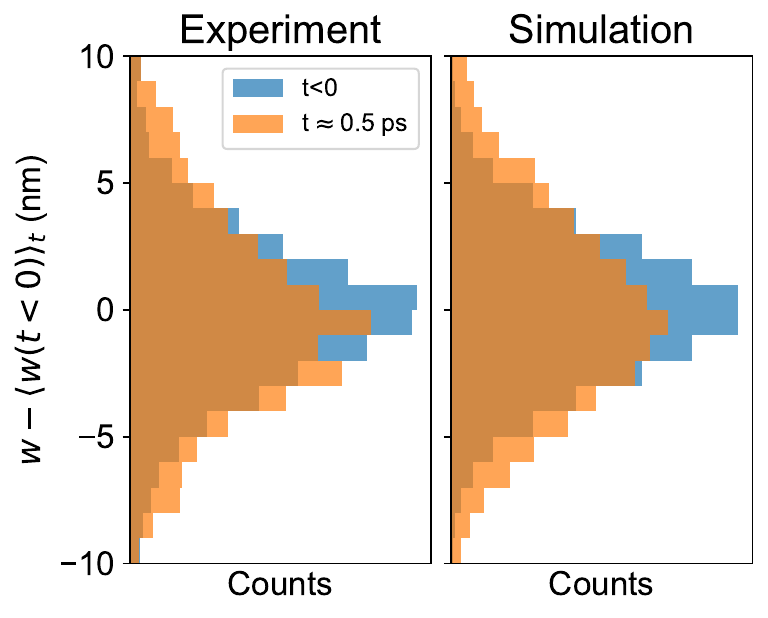}
    \caption{Increase of uncertainty in fitted DW width due to changes of signal-to-noise ratio (SNR). This figure shows a comparison of the empirical changes in width distribution (left panel, cf. Fig. 3c in main text) and the distribution from Monte Carlo simulation with different SNR. Here the ratio between the SNR in the simulations for the two histograms in the right equals the ratio of $M_0$ for the two time delays (left panel).}
    \label{fig:widthMC}
\end{figure}
\begin{figure}
\includegraphics[width=0.4\columnwidth]{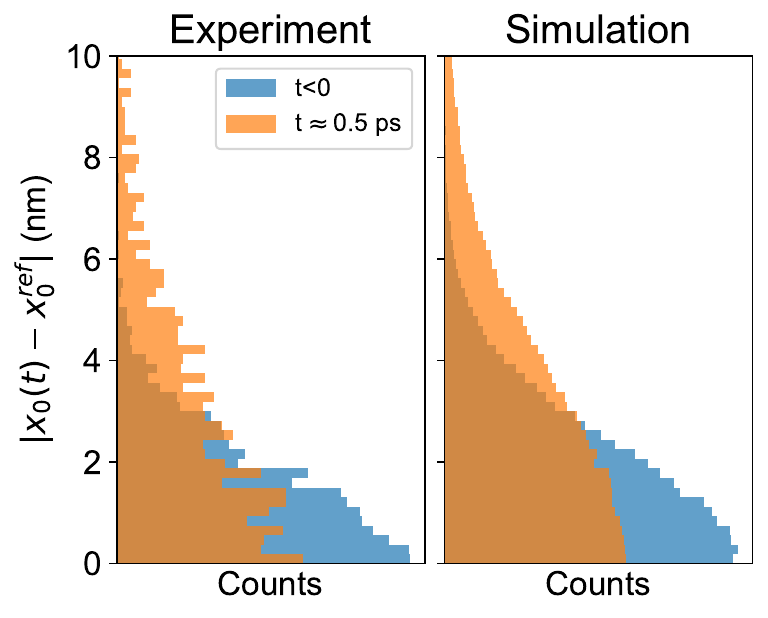} 
\caption {Increase of the DW localization uncertainty for partially demagnetized sample due to a reduction of XMCD contrast. The left panel shows the magnitude of DW displacement (cf. Fig. 3d in main text) and the right panel shows the corresponding results from Monte Carlo simulations with the ratio of SNR equal to the ratio of $M_0$ in experimental data (left panel).}
\label{fig:displacementMC}
\end{figure}

\begin{figure}
\includegraphics[width=\columnwidth]{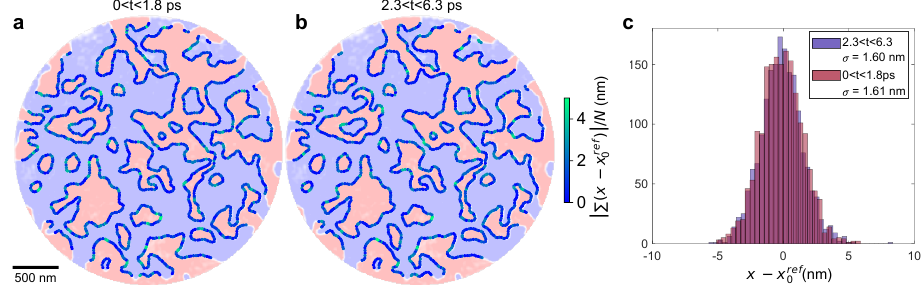}
\caption {Domain wall position stability. \textbf{a} Computed domain wall displacement map for the DW locations identified at the average image of the magnetic state between 0 and 1.8 ps vs positions at the reference point at negative delays. \textbf{b} Local displacement relative to the negative delay, computed for the averaged image of delays between 2.3-6.3 ps. The background is the reference image taken at negative delays.  Dot colors correspond to the magnitude of the local DW displacement of the DW center compared to the positions at a negative delay, $x_{0}^{\mathrm{ref}}$. The corresponding histograms are shown in \textbf{c} for both histograms.}
\label{fig:displacement}
\end{figure}

\begin{figure}
\includegraphics[width=\columnwidth]{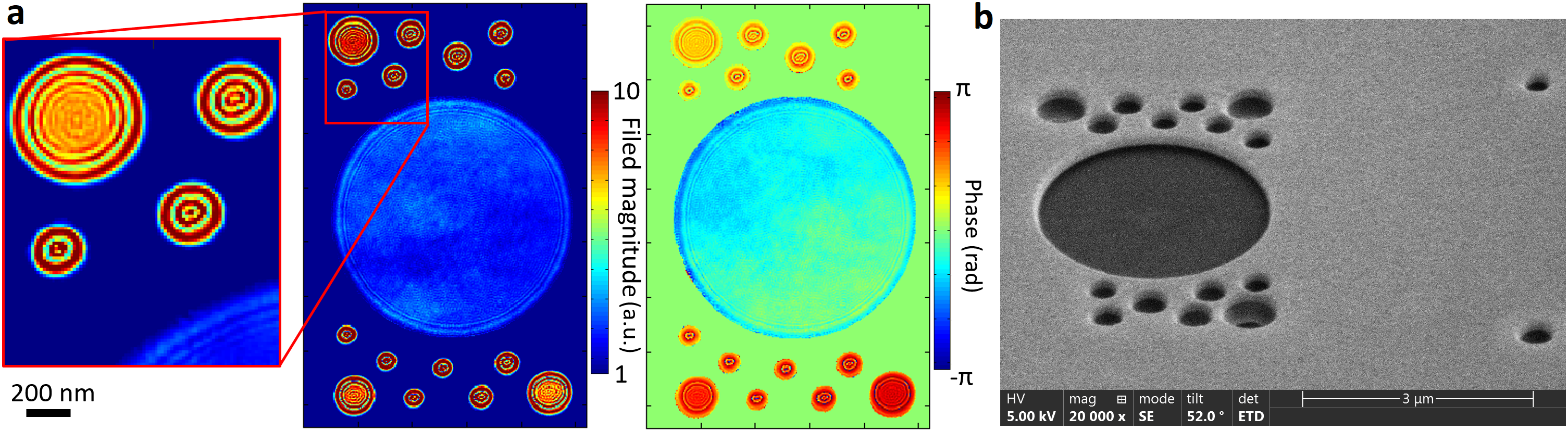}
\caption{\textbf{a} Reconstructed magnitude and phase of the scattered wave after illumination with left-circularly polarized light. The magnified region highlighting the beating of guided modes within the auxiliary holes and at the edge of the FOV aperture. \textbf{b} Tilted view of the mask design captured with scanning electron microscope. The size and arrangement of the auxiliary holes are optimized to ensure a homogeneous far-field diffraction pattern with high scattering angles.}
\label{fig:mask}
\end{figure}

\paragraph*{Resolving local domain wall instability.} Although the distributions of DW width ($w_{avg},w$) and displacement from experimental data stipulate the overall changes in DW properties, it may not be straightforward in reflecting rare events such as DW displacements that only occur at specific locations, e.g. where the DW contour has a large curvature \cite{jangidExtremeDomainWall2023}. In such cases, the derived 0.2~nm precision for the average DW width, for example, may be insufficient to resolve displacements as large as 10~nm if they occur at only a small percentage of DW positions. To accurately assess such DW motions, we focus on local measurements. In addition to the application of the super-Gaussian filter and binning neighboring DW profiles as described in Methods, we further compare results from averaged images from negative delays, delays between 0.1 ps and 1.8~ps (maximal demagnetization), and 2.3 ps-6.3 ps (recovery), with results from individual delays to recover possible displacements at a small number of locations from noise. In Figure~\ref{fig:displacement}, the absolute values of the displacement for averaged images at maximal demagnetization and recovery are plotted with a blue-green colormap in Figs.~\ref{fig:displacement}a,b and the corresponding histograms are shown in Fig.~\ref{fig:displacement}c. In both cases, the standard deviations of the distributions are 1.6~nm, which corresponds to the maximally possible root-mean-square displacement. Note that the value of 1.6~nm is nearly identical to the average of $|x_0-x_0^\mathrm{ref}|$ at negative delays (see main text Fig.~3d). In addition, no correlation is observed between the local curvature of the domain wall and the extracted displacement value (Figs.~\ref{fig:displacement}a,b).
The analysis here further confirms that the extracted changes of DW positions are mainly due to noise, and the upper limit of 3~nm  displacement provided in the main text is a conservative estimation.

\begin{figure}
\includegraphics[width=\columnwidth]{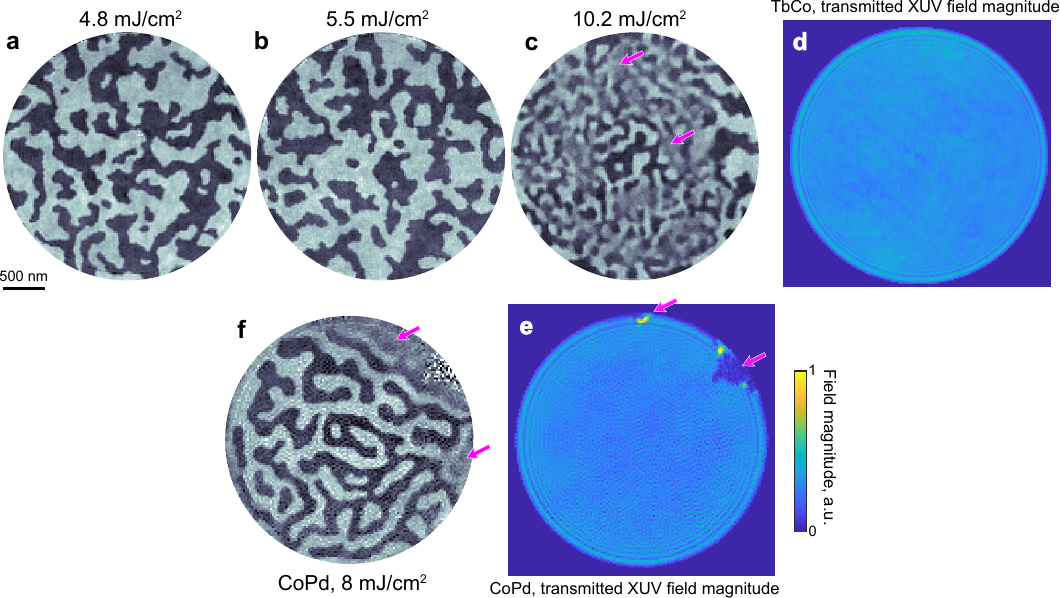}
\caption{\textbf{a–c} Magnetic domain patterns of the TbCo sample recorded at negative delays ($t < 0$) at different FOV locations and pump fluences. \textbf{d} Transmitted field amplitude of the extreme-UV probe beam through the FOV shown in \textbf{c}.
\textbf{e,f} Transmitted field amplitude of the CoPd multilayer sample (e) and the corresponding snapshot of the magnetic domain pattern in this FOV (f). The blurred regions with reduced magnetic contrast, marked by magenta arrows, result from irreversible changes in the domain pattern during the exposure time. Regions with increased XUV transmission and absorption, indicated by magenta arrows in (e), correspond to areas where the magnetic film was damaged by the pump laser.}
\end{figure}

\section{Phenomenological theory of domain wall profiles.} 
Within continuum theory, the domain wall width $w$ is described based on the effective magnetic anisotropy $K$ and exchange stiffness $A$:
\begin{align}
w=\sqrt{\frac{A}{K}}.
\end{align}
In equilibrium, $w$ is temperature dependent due to the magnetization dependence of the micromagnetic parameters. In particular, the seminal theory of Callen et al \cite{Callen1966}, shows that the temperature dependence of the uniaxial anisotropy of a cubic ferromagnet can be expressed as 
\begin{align}
K(T) =K_0\,m(T)^3
\end{align}
where $m(T)=M_s(T)/M_s(0)$ is the reduced magnetization and $M_s(T)$ the saturation magnetization at temperature $T$. Also the exchange stiffness exhibits a temperature dependence. In a mean-field picture, the Landau theory of phase transitions predicts a linear dependence of the exchange stiffness, consistent with a magnetization dependence as follows:
\begin{align}
A(T) =A_0\,m(T)^2.
\end{align}

These phenomenological expressions show that domain walls broaden under the increase of temperature  $\delta(T ) / \delta_0 =m(T)^{-1/2}$. The physics of these temperature dependencies of the micromagnetic parameters can be understood from a coarse graining approach. Assuming temperature-independent single-ion anisotropies and exchange interactions between atomistic spins, effective anisotropies and exchange stiffness for an ensemble of spins are obtained by expanding the free energy for small deviations from the perfectly aligned FM ground state, yielding thermal expectation values $\langle \cos^2\theta \rangle$, where $\theta=0$ corresponds to the FM state. Likewise for exchange stiffness, the temperature dependence follows from $\langle \cos(\theta_{ij}) \rangle$, $\theta_{ij}$ the relative angle between two neighboring spins. 

Numerical calculations confirm the validity of the scaling relations, but generally, different exponents are obtained. In the present case, the thin-film limit is of importance, since for this case, a nearly uniform heating of the system across the film thickness is present, and a non-equilibrium relation between reduced magnetization and micromagnetic parameters may be expected. In this case, the single-ion anisotropy does not provide an adequate description, and an ion-ion anisotropic exchange interaction gives the leading contribution \cite{thiele2002,staunton2004, asselin2010}. This leads to a scaling exponent that is much closer to that of the exchange stiffness:
\begin{align}
K(T) =K_0 m(T)^{2+\kappa},
\end{align}
where $0<\kappa\ll1$. For the exchange stiffness, corrections beyond the mean-field take the form Ref. \cite{atxitia2010}: 
\begin{align}
\quad A(T) =A_0 m(T)^{2-\varepsilon},
\end{align}
where $0<\varepsilon\ll1$. 

Under femtosecond laser excitation, $m(t)$ reduces faster than the magnetic system can reach internal equilibrium. Hence, being coarse-grained parameters, the fastest changes to the domain wall widths are expected when the micromagnetic parameters change in step with the magnetization:
\begin{align}
w[m(t)]/w_0=m(t)^{-(\kappa+\varepsilon)/2}.
\end{align}
Taking literature values $\kappa=0.1$ and $\varepsilon=0.24$, reported for FePt  \cite{thiele2002,atxitia2010}, we obtain $w[m(t)]/w_0 = 1.125$ for a maximum reduction of $m(t)$ by $50\%$. Hence, the maximum estimated domain wall broadening for a 15~nm wall is 1.9~nm, consistent with the experimental error bar. We emphasize that this estimation gives only an upper bound, on short time scales the possible changes are smaller. In particular, on time-scales shorter than $t\ll w_0/v_\mathrm{g}\sim 100$ps, with $v_\mathrm{g}\sim0.1$~nm/ps the highest magnon group velocity, domain wall profiles cannot follow the magnetization instantaneously.

\bibliographystyle{apsrev4-2}
\bibliography{references}